# Who is Funding Indian Research? *A look at major funding sources acknowledged in Indian research papers*

**Vivek Kumar Singh[1,2,3*], Prashasti Singh[4], Anurag Kanaujia[2], Abhirup Nandy[3]**

[1]Department of Computer Science, University of Delhi, Delhi - 110007 (India)
[2]Delhi School of Analytics, University of Delhi, Delhi-110007 (India)
[3]Department of Computer Science, Banaras Hindu University, Varanasi - 221005 (India)
[4]Shri Ram College of Commerce, University of Delhi, Delhi-110007 (India)

**Abstract:** Science and scientific research activities, in addition to the involvement of the researchers, require resources like research infrastructure, materials and reagents, databases and computational tools, journal subscriptions and publication charges etc. In order to meet these requirements, researchers try to attract research funding from different funding sources, both *intramural* and *extramural*. Though some recent reports provide details of the amount of funding provided by different funding agencies in India, it is not known what quantum of research output resulted from such funding. This paper, therefore, attempts to quantify the research output produced with the funding provided by different funding agencies to Indian researchers. The major funding agencies that supported Indian research publications are identified and are further characterized in terms of being national or international, and public or private. The analytical results not only provide a quantitative estimate of funded research from India and the major funding agencies supporting the research, but also discusses the overall context of research funding in India, particularly in the context of upcoming operationalization of Anusandhan National Research Foundation (ANRF).

## Introduction

Science is considered to be one of the greatest collective endeavours, that creates new knowledge. However, science and scientific research activities, in addition to the involvement of the researchers, require resources like research infrastructure, materials and reagents, databases and computational tools, journal subscriptions and publication charges etc. In order to meet these requirements, researchers try to attract research funding from different funding sources, both *intramural* and *extramural*. India has recently emerged as one of the top knowledge producers in the world, with 3rd rank in the global research output in Science and Engineering domain.[1] However, India's gross expenditure on research and development (GERD) has reduced from a high of ~0.90% in 2008 to 0.64% of GDP in 2021, which is quite low when compared to several other countries like USA, China, UK, Germany, South Korea etc.[2] Given the limited funding available to support science and technology research in India, it is important to identify what are the major funding agencies that support research activities in India.

The global expenditure on research and development (R&D) is about 2.5 Trillion USD as per 2022 data, with countries such as the United States and China spending in excess of 500 billion

---

[*] Corresponding Author: vivek@cs.du.ac.in

USD.[3] Competitive research funding is provided by different national agencies, (such as the US's National Science Foundation-NSF, National Institute of Health-NIH, German Research Foundation-DFG, United Kingdom Research and Innovation-UKRI, French National Research Agency-ANR, National Natural Science Foundation of China etc.). Similarly, in case of India the research funding support is provided by various agencies such as Department of Science & Technology (DST), Council for Scientific and Industrial Research (CSIR), University Grants Commission (UGC), Science and Engineering Research Board (erstwhile SERB that is now subsumed in ANRF), Department of Biotechnology (DBT) etc. As per the recent report from DST, respective contributions for research and development comprised central government (43.7%), state governments (6.7%), private sector (36.4%), public sector industry (4.4%) and higher education sector (8.8%).[4] Thus, majority funding to research and development activities is coming from public funding. This is in contrast to patterns of many other developed countries, where the private funding for research and development is in excess of the government funding.[5] In case of India, there has been a long-standing call for increasing the contributions for research from the private sector to also boost the overall GERD to level of above 2%. The ANRF specifically proposes to foster investment in R&D from the private sector entities.[6]

In this context, it is important to understand what is the current status of R&D funding available to Indian research community, identify the major funding agencies, and to quantify the research output produced out of such funding. There lies the objective of this paper. Although, existing reports provide some data about the amount of funding provided by various agencies, the impact of such funding on research outputs is not well documented. Therefore, this paper uses Indian research publication data sourced from Web of Science as a proxy to analyze the quantum of research output resulting out of the research funding provided by the different agencies. More precisely, the study attempts to answer the following research questions:

**RQ1:** What proportion of Indian research output during 2011-22 received a research funding?

**RQ2:** What are the major funding agencies acknowledged in Indian research publications during 2011–22?

**RQ3:** How are these identified funding agencies distributed in terms of being national or international and public or private?

**Related Work**

There have been several previous attempts to analyze the different aspects of research output from India as a whole as well as different institutions and institution systems. For example, many previous studies have tried to analyze India's research output vis-a-vis various other countries.[7-14] The research output in terms of number of research publications of different states of India has also been analysed recently.[15] Similarly, the research outputs of different institutions and institution systems have been analysed in several respects. This includes major centrally funded institutions,[16] research intensive higher education institutions,[17-18] Indian Institutes of Technology (IITs),[19-21] Council of Scientific and Industrial Research (CSIR),[22-23] National Institutes of Technology (NITs),[24-25] Indian Institutes of Science Education and Research (IISERs),[26] ICAR,[27] IIMs,[28-29] Indian Medical Institutions,[30] All India Institute of Medical Sciences (AIIMS),[31] and some selected private universities.[32-33]

The Impact of R&D spending on the levels of R&D undertaken has also been a subject of research by some studies. These include *(a)* India specific studies looking at the spending done by major central government S&T departments on R&D projects,[34-36] health research funding,[37] and *(b)* international outlook specific studies such as international funding patterns among BRICS nations,[38] G9 countries,[39] an Indian perspective on UKRI's policy changes in research funding provided by them,[40] linkages of foreign R&D centres in India[41] etc. Some researchers have used R&D data to comment on the policies and mechanisms of research funding. For example, Hoffman, (2013) in a letter [42] responding to a previously published study, suggested the prerequisite of existing knowledge and experts in a country, above the research funding for higher research productivity, while Manu & Gala, 2020 looked at the research funding data and commented on the data management policies and guidelines of various funding agencies.[43] Some other studies have also explored various dimensions of unfunded research.[44-45]

However, to the best of our knowledge, there are no available studies on identifying the major funding sources acknowledged in the recent Indian research publications, and quantifying the research output produced out of funding provided by different funding agencies to Indian researchers. Therefore, in this work, the major funding agencies that supported Indian research publications are identified and are further characterized in terms of being national or international, and public or private. A large-sized research publication data is analysed for the purpose. The results obtained present interesting insights about research funding in India. Therefore, this study is novel in terms of its objective, analysis, and results.

**Data and Methodology**

The study utilised the research publication data collected from the Web of Science (WoS) database as a proxy of research output. The publication records indexed in the WoS core collection (comprising SCIE, SSCI and AHCI) for the period 2011-22 are obtained and analysed. The following query was formulated to download the data:

(CU="India") AND (PY=2011-2022) AND DT=(Article OR Review)

Only publication records with document type 'article or review' were considered as they represent the primary research output in the database. The analysis methodology is detailed below. The analytical results are computed by writing programs in Python. The complete procedure for analysis is described below and is also shown as a diagram in **Fig. 1** for clarity of understanding and reproducibility.

The search query returned a total of 9,20,284 publication records for India. In order to identify the funded publications from this dataset, the "Funding Orgs" metadata of publication records was processed. A publication record, whose funding information was not available in this field was considered as non-funded, while a publication record whose funding details were found in this field was considered as a funded publication. In this way, a total of 5,04,268 publications were found to be funded. Thereafter, these 5,04,268 publication records were further analysed to identify the funding agencies attributed with the publication records. This was done by processing the "Funding Orgs", "Funding Text" and "Funding Name Preferred" metadata fields of the 5,04,268 publication records. First, all the names of funding agencies were extracted. This was followed by a manual disambiguation of agency names, as different variations of names of the same funding agencies were observed. For example, the Department of Science and Technology (DST) was found to be written as DST, DST India, DST Delhi, etc. in the funding metadata of the publication records. This different variations of names of DST were clubbed as DST and the publication records were tagged accordingly. This procedure was followed for the complete list.

This way we selected a total of 1407 major funding agencies, each of which had at least 50 supported publications during the period. The country of these 1407 funding agencies was identified from the funding metadata of the publication records. In some cases, a web search was also required to correctly identify the funding agency details. The analysis was then performed in two steps. In the first step, the Indian funding agencies were considered and in the next step the foreign funding agencies were considered. This way, a total of 552 funding agencies were identified from India, and the publication records were tagged accordingly. In the second step, the remaining 855 funding agencies belonging to foreign countries were processed in the similar way.

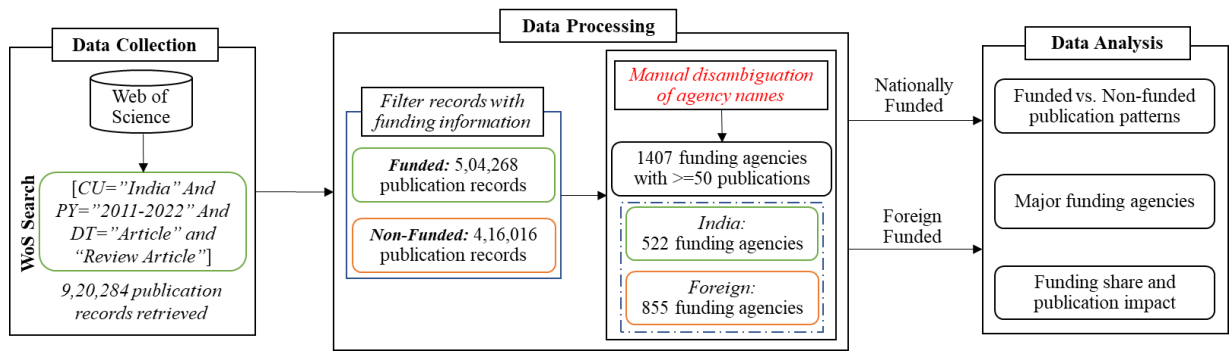

**Fig 1**. Major steps in the analytical methodology

**Results**

*Overall Trends in Funded Publications from India*

The analysis of the Indian research output during 2011-2022 revealed that 54.79% of the 9,20,284 total publications had acknowledged data about the receipt of a funding support, (i.e., Funded TP= 5,04,268) while the rest 45.21% (Non-Funded TP= 4,16,016) did not acknowledge any funding support (**RQ 1**). A comparative overview of the percentage of funded research output of India was also observed vis-a-vis top productive countries like USA, China, Germany, Japan and England during 2011-2022 (**Fig. 2**), for the same document types from WoS database. It was seen that China (~86.48%) accounts for the highest percentage of funded research output followed by Japan (~71.13%) while USA, Germany and England are found to have ~68.01%, 68.59% and ~68.75% of their research output as funded.

The year-wise data of funded and non-funded publications during the period is shown in **Table 1**. A gradual increase is observed in the total share of funded publications over the years with approximately half of the publications acknowledging funding support from the year 2011 itself. The highest percentage of funded publications, about 57% is observed during the years 2015-2018. However, dips in the percentage of funded publications are observed post 2018 which reaches up to 52.68% in the year 2022. In terms of growth rate during 2011-2022, it was observed that the funded research output of India grew with a CAGR (Compounded Annual Growth Rate) of 9.83% while the overall research output of India grew with a CAGR of 9.30%. Thus, the growth of funded research output of India was seen to match pace with the total research output.

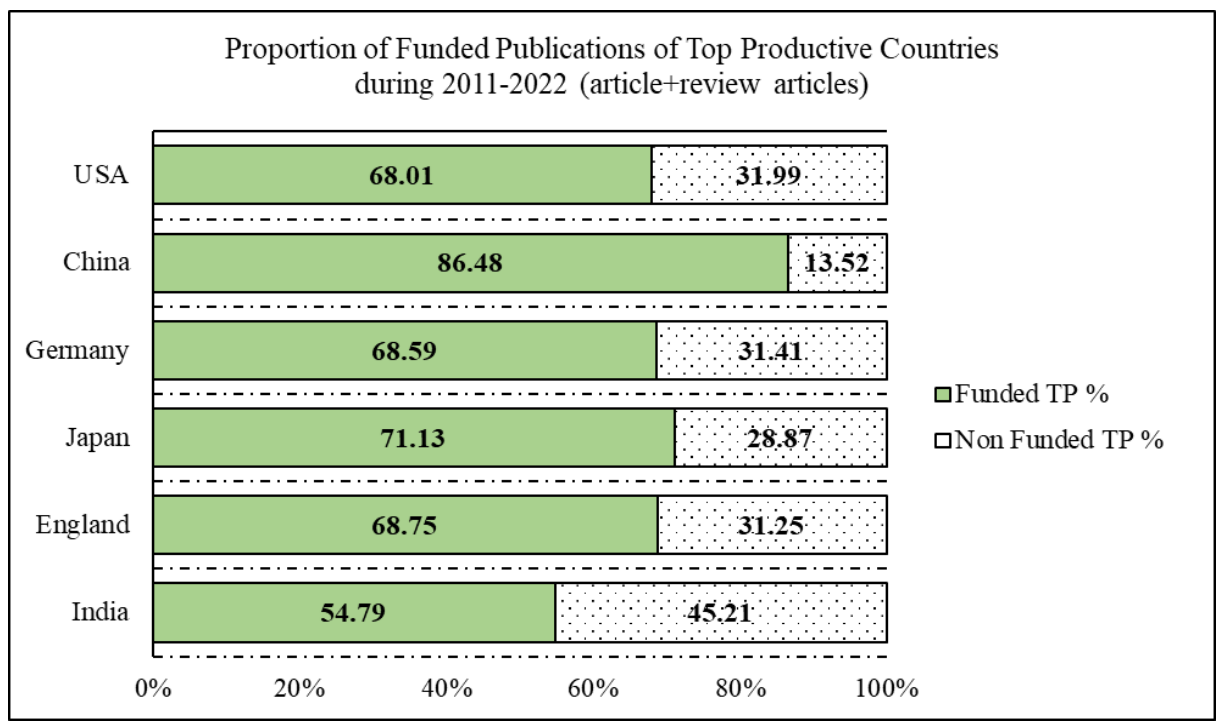

**Fig. 2:** Proportion of Funded and Non-funded Publications of Top Productive Countries during 2011-2022

**Table 1**: Year-wise count and % share of Funded Publications for India

| Year | Total Publications (TP) | Funded TP | % share of Funded TP to TP |
|---|---|---|---|
| 2011 | 47,110 | 23,521 | 49.93 |
| 2012 | 49,634 | 25,826 | 52.03 |
| 2013 | 54,630 | 29,093 | 53.25 |
| 2014 | 59,407 | 33,405 | 56.23 |
| 2015 | 62,320 | 35,581 | 57.09 |
| 2016 | 66,832 | 38,120 | 57.04 |
| 2017 | 70,844 | 40,806 | 57.6 |
| 2018 | 76,284 | 44,114 | 57.83 |
| 2019 | 87,864 | 48,863 | 55.61 |
| 2020 | 1,01,305 | 54,221 | 53.52 |
| 2021 | 1,18,811 | 64,740 | 54.49 |
| 2022 | 1,25,243 | 65,978 | 52.68 |
| **Total** | **9,20,284** | **5,04,268** | **54.79** |

**Note:** CAGR (TP) =9.30%, CAGR (Funded TP) =9.83%

*Major Funding Agencies acknowledged in Indian Research Publications*

The analysis of funded research publications helps identifying the major funding agencies. The top 50 funding agencies identified are listed in **Table 2** along with the supported publication counts. Among the Indian funding agencies, the largest share of funded research output was found to be of DST (22.44%). The CSIR followed next with a share of 16.32%. After DST and CSIR; funding agencies like UGC (14.43%), SERB (8.28%) and DBT (6.23%) were found to have significant count of supported research publications (**RQ 2**). Besides the government funding agencies like DST, CSIR, UGC, SERB and DST; 4 IITs (Kharagpur, Bombay, Madras and Kanpur) along with University of Delhi, IISc, Bangalore and VIT Vellore (a private university) are also found in the list, possibly indicating cases of intra-mural research funding. One may observe around 1000 publications acknowledging the support of the Kerala State Council for Science, Technology and Environment (KSCSTE), the only state funding agency that features in the top agencies list.

**Table 2**: Top Agencies acknowledged in the publications with the count of funded publications during the period (2011-22)

| No. | Funding Agency | Indian/ Foreign | Funded Publications | % share to Funded TP |
|---|---|---|---|---|
| 1 | DST | I | 113174 | 22.44 |
| 2 | CSIR | I | 82286 | 16.32 |
| 3 | UGC | I | 72767 | 14.43 |
| 4 | SERB | I | 41745 | 8.28 |
| 5 | DBT | I | 31404 | 6.23 |
| 6 | DAE | I | 19931 | 3.95 |
| 7 | NIH | F | 18457 | 3.66 |
| 8 | UKRI | F | 17921 | 3.55 |
| 9 | NSF | F | 14043 | 2.78 |
| 10 | ICMR | I | 14014 | 2.78 |
| 11 | National Natural Science Foundation Of China NSFC | F | 9292 | 1.84 |
| 12 | Spanish Government | F | 8741 | 1.73 |
| 13 | JSPS | F | 8580 | 1.7 |
| 14 | MOE | I | 8516 | 1.69 |
| 15 | ICAR | I | 6447 | 1.28 |
| 16 | National Research Foundation Of Korea | F | 6168 | 1.22 |
| 17 | European Union EU | F | 6151 | 1.22 |
| 18 | United States Department Of Energy DOE | F | 6091 | 1.21 |
| 19 | German Research Foundation DFG | F | 5853 | 1.16 |
| 20 | Ministry Of Education Culture Sports Science And Technology Japan MEXT | F | 5368 | 1.06 |
| 21 | DRDO | I | 5099 | 1.01 |
| 22 | King Saud University | F | 4989 | 0.99 |
| 23 | Department Of Science Technology DOST Philippines | F | 4796 | 0.95 |
| 24 | European Research Council ERC | F | 3827 | 0.76 |
| 25 | Conselho Nacional De Desenvolvimento Cientifico E Tecnologico CNPQ | F | 3620 | 0.72 |
| 26 | MOES | I | 3188 | 0.63 |
| 27 | Federal Ministry Of Education Research BMBF | F | 3101 | 0.61 |
| 28 | Chinese Academy Of Sciences | F | 2893 | 0.57 |
| 29 | Fundacao Para A Ciencia E A Tecnologia FCT | F | 2727 | 0.54 |
| 30 | Alexander Von Humboldt Foundation | F | 2694 | 0.53 |
| 31 | Ministry Of Science And Technology Taiwan | F | 2684 | 0.53 |
| 32 | Consultative Group on International Agricultural Research - CGIAR | F | 2680 | 0.53 |
| 33 | Russian Foundation For Basic Research RFBR | F | 2632 | 0.52 |
| 34 | Natural Sciences And Engineering Research Council Of Canada NSERC | F | 2630 | 0.52 |
| 35 | WELLCOME TRUST | F | 2611 | 0.52 |
| 36 | Fundacao De Amparo A Pesquisa Do Estado De Sao Paulo FAPESP | F | 2523 | 0.5 |
| 37 | MEITY | I | 2400 | 0.48 |
| 38 | AICTE | I | 2155 | 0.43 |
| 39 | IIT Kharagpur | I | 1712 | 0.34 |
| 40 | IIT Bombay | I | 1624 | 0.32 |
| 41 | University of Delhi | I | 1572 | 0.31 |
| 42 | IIT Madras | I | 1444 | 0.29 |
| 43 | IISc | I | 1425 | 0.28 |
| 44 | INSA | I | 1327 | 0.26 |
| 45 | ISRO | I | 1061 | 0.21 |
| 46 | IIT Kanpur | I | 1025 | 0.2 |
| 47 | DOS | I | 1004 | 0.2 |
| 48 | MNRE | I | 1003 | 0.2 |
| 49 | KSCSTE | I | 959 | 0.19 |
| 50 | VIT Vellore | I | 953 | 0.19 |

Indian research publications during the period have also received support from various foreign funding agencies as well. In terms of analysing the support provided by foreign funding agencies to Indian research publications during 2011-2022, it was observed that NIH (National Institute of Health, US) supported 18,457 publications, followed by UKRI (UK Research and Innovation department, 17,921 publications), and the NSF (US National Science Foundation, 14,043 publications). Agencies from China, Spain, Japan, Korea, and Germany were also found to have featured among the top 20 funding agencies acknowledged in Indian research output **(RQ2)**. The foreign funding support to these Indian research publications may have been possible through collaborative research schemes and projects. An analysis of the full data of funded publications shows that there are 4,41,988 publications funded by Indian agencies (which is 87.63% of total funded publications) and 3,71,163 publications funded by foreign agencies (which is 73.60% of total funded publications). This suggests that there are a significant number of funded publications that attracted funding support from Indian and foreign agencies. Many of such publication records may be instances of bilateral or multilateral cooperation programs of funding **(RQ3).**

Now we try to analyse the type of funding agencies in terms of being public or private. It can be observed that a major portion of research publications is supported by government agencies, both national and international. There are also instances of some private agencies, research foundations and trusts which were acknowledged. However, the supported publication counts for these agencies is quite low and as a result they do not figure in the top 50 funding agencies acknowledged in Indian research publications. Among these type of agencies from India, the major ones include Hyderabad eye research foundation (510 publications), Tata trusts (325 publications), Infosys foundation (301 publications), Sanofi Aventis (230 publications), GVK Biosciences (50 publications) etc. Among such agencies from foreign countries, the major ones with funded publication counts are Consultative Group on International Agricultural Research – CGIAR (2680), Bill and Melinda Gates Foundation – BMGF (2262), Alfred P. Sloan Foundation (1403), Harrison Goddard Foote Germany (952), The Welch Foundation (892), GlaxoSmithKline (558), Novartis (450), Johnson & Johnson (438), AstraZeneca (438), Pfizer (435), Merk (363), Bristol Myers Squibb Biotech (342), NVIDIA Corporation (323), Eli Lilly pharmaceutical (313), Kavli Foundation (308), Simons foundation (308), Carlsberg Foundation (282), Roche (191), Abbott (143), Bayer (171) etc. **(RQ3)**. It may be noted that some of the foundations and agencies such as CGIAR act as platforms for consolidating funds from national governments, private foundations, and multilateral funding and development agencies. These observations show that among the private agencies, research foundations and trusts, more foreign entities are acknowledged. However, the amount of government funding far exceeds and indicates the dominant role played by this type of funding in the national research activities.

**Discussion**

This study has analysed research publications from India during 2011-22 to identify the major funding sources acknowledged. It computed the year wise patterns and growth of Indian research publications and the funded Indian research publications. The proportionate share of funded publications of India is compared with patterns of some other countries namely, USA, China, Germany, Japan, and England. It is observed that the proportion of funded research publications from India ranged between 49-58% with a cumulative of 5,04,268 funded publications (54.79%) **(RQ1)**. This level is however much lower than the other countries considered, such as of China (86.48%), US (68.01%) etc. This indirectly indicates about availability of more research funding opportunities in these countries. When seen in the light of the Gross Expenditure on Research and Development (GERD) by these countries, it becomes evident that India needs to invest much more in R&D. Further, many of these countries have a good part of research funding from non-government sources, indicating the need for exploring opportunities for research from various non-government sources in India.

The analysis identifies the major funding sources, based on the acknowledgement written by authors in the research publications. The most acknowledged agencies in Indian research publications are DST, CSIR, UGC, SERB, and DBT. Among the foreign funding agencies, the major ones include NIH, UKRI, NSF, NSFC etc. **(RQ2)**. A small set of the research publications have acknowledged private agencies, research foundations and trusts. Some major agencies providing such support include Hyderabad eye research foundation, Tata trusts, Infosys foundation, Sanofi Aventis, GVK Biosciences, BMGF, Alfred P. Sloan Foundation, The Welch Foundation, GlaxoSmithKline, NVIDIA etc. Among these, more foreign entities are acknowledged than Indian entities, underlining the lack of enough private funding sources in India **(RQ3)**.

In this context, it may be relevant to correlate the findings of the present study with data on quantum of funding by various agencies as provided by the R&D statistics in the DST report.[4]

DST (funding up to INR 1,402.8 Crore and named in 22% publications), DBT (funding up to INR 330.8 Crore and named in 6.23% publications), Ministry of Electronics and IT (MeitY, funding up to INR 310.6 Crore and named in 0.48% publications), Ministry of New and Renewable Energy (MNRE, funding up to INR 221.5 crores and named in 0.2% publications), and the Defence Research and Development Organisation (DRDO, funding up to INR 70.9 crores and named in 1.01% publications), are the top five bodies providing extramural funding support R&D activities during the year 2020–21. It may be noted that DST's R&D expenditure based was in excess of INR 3,000 crore out of which, INR 1,402.8 crore is spent on Extramural research (EMR) grants by the agency. The EMR funding by DST is highest among the Indian funding agencies which also corresponds to the highest number of funded publications. On the other hand, the funding support from private sector is limited and well below global trends for private investment in R&D at 65%.[4] This may be the reason for a significantly lower number of research publications acknowledging these agencies.

In continuation to the discussion above, one may also like to look at the operationalisation of the ANRF, which along with other roles, aims to: (a) assist in setting up conducive research infrastructure and environment, and (b) support research into capital intensive technology developments, by (c) encouraging public sector enterprises and private sector entities to invest in its activities.[6] The focus of ANRF leans upon the private sector investment heavily, proposing about two-thirds (~70%) of the overall funding from the private entities while the rest one-third (~30%) comes from the government.[46] Bringing together the public and private agencies for supporting research into capital intensive technologies is going to be a challenging task to be taken up by ANRF. It is more relevant to see this in the context of the findings of this study which tell us that the major portion of funded research publications from India are supported by government funding, both among Indian as well as foreign funding agencies. Only a few foreign funding agencies were private entities and even less among the Indian entities are private. These observations thus show a completely different pattern than what is envisaged by the ANRF, and therefore it will need a complete paradigm shift for ANRF to be able to change the funding patterns for Indian research ecosystem.

*Limitations*

The data for this study was retrieved from Web of Science database which does not index all the research publications[47] as a result there may be other research publications resulting out of different funding but not captured in the data of the present study. Another point to remember here is that the funding agency information is retrieved from the publications from the relevant metadata. In case of publications where the authors of the publication fail to acknowledge funding support or the funding metadata is missing for some other reason, the publication is not counted as an instance of funded publication even if it may have received funding support. Furthermore, the study only considers research publications resulting out of research funding, whereas such funding may result into different kinds of outcomes such as patents, technological developments, R&D infrastructure, and capacity building etc. These outcomes, subject to availability of data, may also be considered in future exercises.

**Conclusion**

To study analysed the research publications from India during the period 2011-22 with the purpose of quantifying the research output produced out of funding provided by different funding agencies to Indian researchers. Further, the major funding agencies that supported Indian research publications are identified and are further characterised in terms of being national or international, and public or private. The proportionate share of funded research publications in India is found to be significantly lower as compared to other major knowledge producing countries. Further, majority of the funded research output is supported by public funding agencies with very low funded research output supported by private agencies and non-government organizations. The results highlight the need for increased research funding in India and also suggests for the need to widen the research funding support ecosystem by including various non-government agencies. The ongoing process of operationalization of Anusandhan National Research Foundation (ANRF) may be a major enabler in this direction and the Indian research community is eagerly awaiting these positive changes.